\documentclass[aps,superscriptaddress,nofootinbib,preprint,tightenlines,showpacs,eqsecnum]{revtex4}
\usepackage{hyperref}
\usepackage{epsfig,rotating}
\usepackage{amsmath,amssymb}
\usepackage{dsfont}

\newcommand{\pslash}{\not\!{p}}

\newcommand{\be}{\begin{equation}}
\newcommand{\ee}{\end{equation}}
\newcommand{\bea}{\begin{eqnarray}}
\newcommand{\eea}{\end{eqnarray}}

\begin{document}

\title{Oscillation dynamics of active-unsterile neutrino mixing \\
in a  $2+\tilde{1}$ mixing scheme}
\author{D. Boyanovsky}
\email{boyan@pitt.edu} \affiliation{Department of Physics and
Astronomy, University of Pittsburgh, Pittsburgh, PA 15260}
\author{R. Holman} \email{rh4a@andrew.cmu.edu}
\affiliation{Department of Physics, Carnegie Mellon University,
Pittsburgh, PA 15213}
\author{Jimmy A. Hutasoit}\email{jhutasoi@andrew.cmu.edu}  \affiliation{Department of Physics, Carnegie Mellon University, Pittsburgh, PA 15213}

\date{\today}

\begin{abstract}
We consider the possibility that sterile neutrinos exist and admit a
description as unparticles; we call these {\em unsterile} neutrinos.
We then examine the nature of neutrino oscillations in a   model
 where an unsterile can mix with two active flavors with a
 very simple  mass matrix of the see-saw type. Despite these simplifications,
  we find a number of remarkable features, all of which will occur generically when more realistic
   cases are considered. These include momentum dependent mixing angles, ``invisible''
   decay widths for the unsterile-like mode, as well as the inheritance of a non-vanishing
   spectral density for the massive active-like modes. We also obtain the disappearance and appearance
   probabilities for the  active-like neutrinos and find remarkable interference effects between the active
   and unsterile neutrinos as a consequence of threshold effects, yielding new oscillatory contributions with
   different oscillation lengths.  These effects are only measurable on short baseline experiments because there both probabilities are suppressed as compared to mixing with a canonical sterile neutrino, thereby providing a diagnostics tool to
   discriminate unsterile from canonical sterile neutrinos. We conclude with a discussion of
   whether these new phenomena could aid in the reconciliation of the LSND and MiniBooNE results.
\end{abstract}


\maketitle

\section{Introduction}
Neutrino masses and oscillations are now an established phenomenon and indisputable evidence of physics beyond the standard model. Although the origin and scale of  these masses remains a challenging question, the see-saw mechanism provides a compelling framework on how to explain the small active neutrino masses \cite{seesaw}.

One particular extension of the Standard Model consists of adding the so called sterile neutrinos, which are massive $SU(2)$ singlets. Scenarios involving sterile neutrinos with mass in the $\mathrm{GeV}$ range have been proposed as explanations of two seemingly unrelated and unexpected phenomena: an excess of air shower events at the SHALON gamma ray telescope \cite{shalon}, and the prominent peak of electron-neutrino events  above background for $300 \, \mathrm{MeV} \leq E_\nu \leq 475\, \mathrm{MeV}$ at  the MiniBooNE \cite{miniboone}. Examples of such scenarios include a  3+2 oscillation scheme involving three active neutrinos and two sterile neutrinos proposed in Ref. \cite{maltoni}, and  an alternative explanation invoking the radiative decay of the heavy sterile neutrino with a small magnetic moment  proposed in Ref. \cite{gninenko}.

Recently, Georgi \cite{georgi} suggested another extension of the Standard Model, which includes a ``hidden" conformal invariant sector with a non-trivial infrared fixed point. This conformal sector can be realized by a renormalization flow toward the infrared through dimensional transmutation \cite{banks, fox}. Below the dimensional transmutation scale, there emerges an effective interpolating field, the unparticle field, that features an anomalous scaling dimension. Various studies recognized important phenomenological
 \cite{georgi,cheung,spin12,ira,Liao:2007ic,zwicky1,flavor,parity,chiumanho}, astrophysical \cite{raffelt,deshpande,freitas} and cosmological \cite{macdonald,davo,lewis,kame,he,kiku,xuelei,rich,unparticleus} consequences of unparticles. The consequences of the existence of scalar unparticles in the neutrino sector have also been studied. A scenario where  the heaviest mass eigenstates corresponding to the mixed $\nu_\mu-\nu_e$ decaying into the lightest eigenstate and a scalar unparticle has been  suggested as an explanation of the MiniBooNE anomaly \cite{liliu,Zhou:2007zq}. Furthermore, unparticle contributions to the neutrino-nucleon cross section and its influence on the neutrino flux expected in a neutrino telescope such as IceCube have been reported in Ref. \cite{telescopes}.

The question arises whether interesting physics might be found by combining these two ideas. Thus, suppose that sterile neutrinos {\em do} exist, and furthermore, they admit a description in terms of unparticles. What consequences arise for neutrino physics, and in particular, neutrino oscillations?

In Ref. \cite{Boyanovsky:2009ke}, we considered the sterile neutrino
to be an unparticle, namely an \emph{unsterile neutrino}, and
studied the mixing of an active and an unsterile neutrino, which
revealed unexpected momentum dependence of the mixing angles. Here,
we extend that model by including a second active neutrino, and
study the consequences in neutrino oscillations. To be precise, we
consider a $2+\tilde{1}$ scheme in which two massless active
neutrinos mix with an unsterile neutrino via a simple mass matrix of
the see-saw type that mixes the active neutrinos only with the
unsterile one. Despite the simplicity of the  mass matrix, a number
of results arise that will persist in more complex models.

We find that the unsterile-active mixing angle depends on the four-momentum and that the propagating modes consist of a massless active-like mode, a massive active-like mode and an unsterile-like mode. The structure of propagators of the massive active-like mode and the unsterile-like mode are similar to the ones found in \cite{Boyanovsky:2009ke}.

To study the dynamics of active-active oscillation, we generalize  the conventional quantum mechanical description of neutrino oscillations  in terms of the propagators of mass eigenstates  and  calculate the appearance and disappearance probabilities. The oscillation dynamics reveals remarkable interference phenomena  associated with the threshold of the inherited spectral density for the massive active-like and the complex unsterile-like pole. The momentum dependent mixing angles along with these novel interference contributions yield new oscillatory contributions to the appearance and disappearance probabilities of the active neutrinos with a different oscillation length. These new contributions will be  manifest as quantum beats as a function of energy and baseline in oscillation experiments. The non-oscillatory contributions to these probabilities are suppressed with respect to the case of a canonical sterile neutrino, which has no anomalous dimension.

Recently, in Ref. \cite{Schwetz:2007cd}, it has been proposed that energy dependent mixing and or oscillation parameters \emph{may} reconcile the LSND \cite{lsnd} and MiniBooNE \cite{miniboone} data. This suggestion hinges on the possibility of ``exotic'' sterile neutrinos whose mixing with the active ones features novel energy dependent mixing angles and/or oscillation parameters.
The energy dependence may reconcile the results of both experiments which are performed in different energy regimes ($\sim 40 \,\mathrm{MeV}$
for LSND, vs. $\sim \,\mathrm{GeV}$ for MiniBooNE). Given the energy dependence we find in our analysis, we were motivated to see whether our model could be of use in understanding how the LSND and MiniBooNE data mesh together.
 For our   model, we find that we cannot reconcile the data, but that might be attributable to not having enough structure in the mass matrix.

\section{$2+\tilde{1}$ Scheme}
The Lagrangian density for the unsterile in momentum space is \cite{Boyanovsky:2009ke}
\be \mathcal{L} = \overline{\psi}_U(-p)
\, \big(\pslash -M\big)F(p)\,\psi_U(p), \label{unlag}\ee
where
\be
F(p) = \Big[\frac{-p^2+M^2-i 0^+}{\Lambda^2}\Big]^{-\eta}~~,~~ 0
\leq \eta < 1 \,.\label{F}\ee
$\Lambda$ is the scale below which the low energy dynamics is dominated by the infrared fixed point of the conformal sector. Below this scale the unparticle is described by an interpolating field whose two point correlation function scales with an anomalous dimension. Consistency of the unparticle interpretation requires that $M < \Lambda$, where $M$ is the infrared cut-off or unparticle threshold. This Lagrangian density can be understood using a renormalization group resummation  argument discussed in Ref. \cite{Boyanovsky:2009ke} and references therein.

We consider a see-saw type mixing with two active Dirac neutrinos, such that the full momentum space Lagrangian density is given by
\be
\mathcal{L} = \left( \begin{array}{ccc} \overline{\nu}_{a_1}(-p) & \overline{\nu}_{a_2}(-p) & \overline{\psi}_U(-p) \end{array}\right)\,\left(\begin{array}{ccc} \pslash & 0 & -m_1\\ 0 & \pslash & -m_2 \\  -m_1 & -m_2 & (\pslash - M) F(p) \end{array} \right)
\,\left(\begin{array}{c}
                   \nu_{a_1}(p) \\
                   \nu_{a_2}(p) \\
                   \psi_U(p)
                 \end{array}\right),
\ee
with $\nu_{a_{1,2}}$ are the active neutrinos and we assume $m_{1,2}\ll M < \Lambda$ leading to a see-saw hierarchy of masses. Following \cite{Boyanovsky:2009ke}, let us introduce a rescaled unsterile field
\be
\nu_U = \sqrt{F(p)}~~ \psi_U, \label{unfieldred}
\ee
such that the Lagrangian density is now given by
\be
\mathcal{L} = \left( \begin{array}{ccc} \overline{\nu}_{a_1} & \overline{\nu}_{a_2} & \overline{\nu}_U \end{array}\right)\, \left(\pslash \, \mathds{I} - \mathds{M}\right) \, \left(\begin{array}{c}
                   \nu_{a_1} \\
                   \nu_{a_2}\\
                   \nu_U
                 \end{array}\right),
\ee where $\mathds{I}$ is the identity in the ``flavor" space, and
the ``mass" matrix is \be \mathds{M} = \left(\begin{array}{ccc} 0 &
0 & \frac{m_1}{\sqrt{F}}\\ 0 & 0 & \frac{m_2}{\sqrt{F}} \\
\frac{m_1}{\sqrt{F}} & \frac{m_2}{\sqrt{F}} & M \end{array} \right).
\label{Mmat} \ee We note that $\mathds{M}$ has a zero eigenvalue,
which aids considerably in the diagonalization. We reiterate that
this mass matrix describes a simplified model, chosen only so as to
exhibit the mixing that we want to describe (active-unsterile)
 while suppressing effects we consider extraneous to the discussion.

This mass matrix can be diagonalized exactly and the fields corresponding to the propagating or ``mass" eigenstates are given by
\bea
\nu_1 &=&  \frac{-m_2\,\nu_{a_1} +m_1\,\nu_{a_2}}{m} \, , \label{massless}\\
\nu_2 &=&  \frac{   2 m_1\, \nu_{a_1} +  2 m_2 \, \nu_{a_2} - \left(\sqrt{M^2 F + 4 m^2} - M \sqrt{F}\right)\, \nu_U}{\left[2 M^2 F +  8 m^2 - 2 M \sqrt{M^2 F^2 + 4m^2 F} \right]^{1/2}} \, , \label{massiveactive}\\
\nu_3 &=&  \frac{2 m_1\, \nu_{a_1} + 2 m_2 \, \nu_{a_2} + \left(M \sqrt{F} + \sqrt{M^2 F + 4 m^2}\right)\, \nu_U}{\left[2 M^2 F +  8 m^2 + 2 M \sqrt{M^2 F^2 + 4 m^2 F} \right]^{1/2}}, \label{unsterile}
\eea
with their propagators given by
\bea
G_1 &=& \frac{\pslash}{p^2} \, , \\
G_2 &=& \frac{\pslash + M_2(p)}{p^2-M_2^2(p)}\, ; \qquad M_2 = \frac{M\sqrt{F} - \sqrt{M^2 F+ 4 m^2}}{2 \sqrt{F}}\, ,\\
G_3 &=& \frac{\pslash + M_3(p)}{p^2-M_3^2(p)}\, ; \qquad M_3 = \frac{M\sqrt{F} + \sqrt{M^2 F+ 4 m^2}}{2 \sqrt{F}}\, ,
\eea
respectively, where  \be m = \sqrt{m_1^2 + m_2^2}\label{mas}\,. \ee
The propagators of the massive active-like mode and unsterile-like mode follow closely the propagators of the active-like mode and unsterile-like mode of Ref. \cite{Boyanovsky:2009ke}.

As in Ref. \cite{Boyanovsky:2009ke}, we assume that $m_{1,2}\ll M$ and self-consistently
\be \frac{m^2}{M^2 F(p_{2,3})} \ll 1 \,.\label{aproxi}\ee In this approximation, it is convenient to introduce \be  \epsilon_{1,2}(p) = \frac{m_{1,2}}{M\,\sqrt{F(p)}}~~;~~ \epsilon(p) = \sqrt{\epsilon^2_1(p)+\epsilon^2_2(p)}\, ,  \label{epsilons} \ee  and the mixing angles
\bea \cos \theta  = \frac{\epsilon_1}{\epsilon}= \frac {m_1}{m} ~~;~~\sin \theta  = \frac{\epsilon_2}{\epsilon} = \frac {m_2}{m} ~~;\nonumber \\ \cos \Phi  = \frac{1}{\sqrt{1+\epsilon^2(p)}} ~~;~~\sin \Phi  = \frac{\epsilon(p)}{\sqrt{1+\epsilon^2(p)}} \,. \label{mixangles}\eea  We note that whereas the angle $\theta$ does not depend on momentum,   the angle $\Phi$ \emph{does} depend on momentum through $F(p)$.

To leading order in $\epsilon$'s we find \be M_2(p)= - M \epsilon^2(p)~~;~~M_3(p) = M\big[1+\epsilon^2(p)\big]\,, \label{bigmasses}\ee and   the fields associated with the propagating eigenstates given  by eqs. (\ref{massless}) - (\ref{unsterile}) can be written as
\bea
\nu_1 &=& - \sin \theta \, \nu_{a_1} + \cos \theta \, \nu_{a_2}\,, \label{0angle} \\
\nu_2 &=& \cos \Phi \, \cos \theta \, \nu_{a_1} + \cos \Phi \, \sin \theta \, \nu_{a_2} - \sin \Phi \, \nu_U \, ,   \label{lightangle}\\
\nu_3 &=& \sin \Phi \, \cos \theta \, \nu_{a_1} + \sin \Phi \, \sin \theta \, \nu_{a_2} + \cos \Phi \, \nu_U \, . \label{heavyangle}
\eea

The properties of the propagators can be found by following the calculation in Ref. \cite{Boyanovsky:2009ke} and in the following, we will summarize the results. For details, see \cite{Boyanovsky:2009ke}.

\subsection{  Poles and Spectral Density for the Massive Active-like Mode}
Near the pole, the propagator behaves as 
\be
\frac{1}{p^2 - M^2_2(p)} \approx \frac{Z_2}{p^2 - {\cal M}^2_2}\,, \label{alfaplusbeta}
\ee
where
\be
Z^{-1}_2 \approx   1+ 2\eta ~ \frac{{\cal M}^2_2}{M^2}  \,, \label{Z1}
\ee
and
\be
{\cal M}^2_2 = \left( \frac{M\Delta}{2}\right)^2 \,, \label{p2min}
\ee
with
\be \Delta = 2~\frac{m^2}{M^2} \Bigg[\frac{M^2}{\Lambda^2} \Bigg]^{\eta}\,.\label{dimvars}\ee
This is an isolated pole  below  the unparticle threshold at $p^2 = M^2$. This pole lies on  the real $p^2$ axis and describes a massive stable active-like propagating mode.

The massive active-like propagator also features an inherited spectral density
\be
\rho_2(x) = \frac{\Theta(x)}{\pi} ~\frac{\frac{\Delta^2}{4}\,x^{2\eta}\sin(2\pi\eta)} {\Big[ x+1 - \frac{\Delta^2}{4}\,x^{2\eta}\cos(2\pi\eta) \Big]^2+\Big[\frac{\Delta^2}{4}\,x^{2\eta}\sin(2\pi\eta) \Big]^2}\, .  \label{disc1}
\ee
where
\be
x = \frac{p^2-M^2}{M^2} \,. \label{dimvarx}
\ee
The non-vanishing spectral density above the unparticle threshold at $p^2 = M^2$ will lead to interesting new phenomena in the disappearance and appearance probabilities discussed below.

\subsection{Complex Poles and Spectral Density for the Unsterile-like Mode}
The solution describes a pole in the complex plane (a resonance) and it exists \emph{only} for $$\mathrm{Re}(x) > 0~,~ 0 \leq \eta < 1/3.$$
Near this pole we find
\be
\frac{1}{p^2-M^2_3(p)} \approx \frac{Z_3}{p^2-{\cal M}^2_3+i {\cal M}_3\Gamma}\,, \label{alfaminbeta}
\ee
with
\be
Z_3 = \frac{1}{1-\eta}\,, \label{Z2}
\ee
\be
{\cal M}^2_3 = M^2 \Big[1 + \Delta^\frac{1}{1-\eta}\, \cos\Big( \frac{\pi \eta}{1-\eta}\Big)\Big]\;,\label{M22}
\ee
and
\be
\Gamma =\frac{ M^2}{{\cal M}_3}   \Delta^\frac{1}{1-\eta}\,\sin\Big( \frac{\pi \eta}{1-\eta}\Big) \;. \label{gama}
\ee
The imaginary part is a consequence of the fact that the real part of the pole is above the unparticle continuum and it describes the \emph{decay} of the unsterile like mode into the massive active-like mode and particles in the ``hidden'' conformal sector. This can be understood using a renormalization group resummation argument presented in Ref. \cite{Boyanovsky:2009ke}.
The spectral density is obtained from the discontinuity across the real axis in the complex $p^2$ plane
\be
\rho_3(x) = \frac{\Theta(x)}{\pi} ~\frac{ \Delta \,x^{\eta}\sin(\pi\eta)}{\Big[ x  - {\Delta }\,x^{\eta}\cos(\pi\eta) \Big]^2+\Big[{\Delta}\,x^{\eta}\sin(\pi\eta) \Big]^2}\,.  \label{disc2}
\ee

For consistency of the see-saw mechanism, to give small masses to the active-like neutrinos it is required that \be \Delta  =  2~\frac{m^2}{M^2} \Bigg[\frac{M^2}{\Lambda^2} \Bigg]^{\eta} \ll 1 \,, \label{smalldelt}\ee thus ensuring that $\mathcal{M}_2= \frac{1}{2} M \Delta \ll M$.

\section{Towards Understanding Active-Active Oscillation Dynamics.}

The familiar quantum mechanical description of neutrino oscillation invokes single particle quantum mechanical states. The essential ingredients are the expectation values
$\langle \nu_i | \, e^{-iHt}\,|\nu_i\rangle $. In quantum field
theory, the (fermionic) fields are expanded in terms of creation and
annihilation operators of quanta associated with a Fock basis, which
is determined by the spinor basis functions. For the mass
eigenstates these are the solutions of the (free) Dirac equation,
whereas for flavor states such choice is ambiguous
\cite{boyahodense}. To focus the discussion on a simple case to be
generalized later, consider a flavor doublet of Dirac neutrinos with
an off-diagonal mass term, with Lagrangian density \be \mathcal{L} =
\overline{\Psi}\Big[ \pslash -\mathds{M} \Big]\Psi
\,,\label{lag2flav}\ee where \be \Psi = \left(
                                   \begin{array}{c}
                                     \psi_e \\
                                     \psi_\mu \\
                                   \end{array}
                                 \right) ~~;~~\mathds{M} = \left(
                                                             \begin{array}{cc}
                                                               m_{ee} & m_{e\mu} \\
                                                               m_{e\mu} & m_{ \mu\mu} \\
                                                             \end{array}
                                                           \right).
\label{flavspimass}\ee Diagonalizing the mass matrix via a unitary transformation leads to the fields $\psi_{1,2}$ associated with the mass
eigenstates $m_{1,2}$, related to the flavor fields $\psi_{e,\mu}$ as \bea \psi_e &=& \cos \theta \,\psi_1 + \sin \theta \,\psi_2\,,
\label{psie}\\ \psi_\mu &=& \cos \theta \,\psi_2 - \sin \theta \,\psi_1 \,. \label{psimu}\eea  The neutrino fields $\psi_{1,2}$ are
quantized as usual\footnote{We keep the same notation for the field and its spatial Fourier transform to simplify notation, no
confusion should arise since the arguments are different.} \bea \psi_i(\vec{x},0) & = &  \ \frac{1}{\sqrt{V}} \sum_{\vec{k},h=\pm 1}
e^{i \vec{k}\cdot\vec{x}}\, \psi_i(\vec{k},0)\,,\nonumber \\ \psi_i(\vec{k},0) & = &
\Big[b_i(\vec{k},h) U_i(\vec{k},h) + d^\dagger_i(\vec{k},h) V_i(-\vec{k},h)\Big] ~~;~~i=1,2 \label{psiqua}\eea
where the annihilation operators $b_i,d_i$ and creation
operators $b^\dagger_i,d^\dagger_i$ obey the usual anticommutation
relations and the spinors $U,V$ are solutions of the Dirac equations
\bea \big[\vec{\alpha}\cdot \vec{k} + \beta m_i] U_i(\vec{k},h) & =
& E_i(k) U_i(\vec{k},h)\,, \label{Us}\\  \big[\vec{\alpha}\cdot \vec{k}
+ \beta m_i] V_i(-\vec{k},h) & = & -E_i(k) V_i(-\vec{k},h)\,,\label{Vs}
\eea with $E_i(k) = \sqrt{k^2+m^2_i}$, and are eigenstates of the
helicity operator with eigenvalues $h=\pm1$. Obviously there is no
\emph{unambiguous} quantization of the \emph{flavor spinor fields }
because there is no \emph{unambiguous} choice of basis spinors $U,V$
for these \cite{boyahodense}. One \emph{could} chose spinor solutions of the massless
Dirac equation or with the masses in the diagonal entries in the
mass matrix $\mathds{M}$. Any of these choices correspond to spinor solutions
which are not orthogonal to any of the solutions of the mass
eigenstates (either positive or negative energy) \cite{boyahodense}.

The positive energy
\emph{single particle} quantum mass eigenstates of helicity $h$ are \bea |\nu_i(\vec{k},h)\rangle = b^\dagger_i(\vec{k},h) \,|0\rangle = \psi^\dagger_i(\vec{k},0) \,|0\rangle \,U_i(\vec{k},h)\, , \label{bra} \\ \langle \nu_i(\vec{k},h)| = \langle 0 | \,  b_i(\vec{k},h) =  U^\dagger_i(\vec{k},h) \langle 0 | \, \psi_i(\vec{k},0)\,. \label{ket}\eea
We \emph{define} the single particle flavor states as \bea |\nu_e \rangle & = & \cos \theta \, |\nu_1\rangle  + \sin \theta \, |\nu_2 \rangle\,,  \label{nuesta}\\
|\nu_\mu \rangle & = & \cos \theta \, |\nu_2\rangle  - \sin \theta
\, |\nu_1 \rangle  \,, \label{numusta} \eea where the quantum numbers $\vec{k},h$ are common to all the states. Since the Hamiltonian is diagonal in the mass basis, the transition amplitudes
$\langle \nu_\alpha|\, e^{-iHt}\,|\nu_\beta\rangle$ ($\alpha,\beta =
e,\mu$) require the overlaps \be  \langle \nu_i(\vec{k},h)|\,
e^{-iHt}\, |\nu_i(\vec{k},h)\rangle = U^\dagger_i(\vec{k},h) \langle
0|\,\psi_i(\vec{k},t) \overline{\psi}_i(\vec{k},0)\,|0 \rangle
\,\gamma^0  U_i(\vec{k},h)\,,\label{overlapt}\ee where
$\psi_i(\vec{k},t)$ are the Heisenberg field operators. In terms
of the propagator for $t>0$, the overlaps are given by
\be \langle \nu_i(\vec{k},h)|\, e^{-iHt}\, |\nu_i(\vec{k},h)\rangle =
 U^\dagger_i(\vec{k},h)\,\big( i S_i(\vec{k},t>0)\big) \gamma^0 U_i(\vec{k},h) \,,\label{propover}\ee
 thereby establishing a direct relation between the quantum field theory propagators for the mass eigenstates and the single particle transition amplitudes.

It is convenient to work in the chiral representation \be\gamma^0  = \left(
                                                          \begin{array}{cc}
                                                            0 & -\mathds{I} \\
                                                            -\mathds{I} & 0 \\  \end{array}
                                                        \right)~~;~~ \vec{\gamma} =
\left(
\begin{array}{cc}
  0 & \vec{\sigma} \\
  -\vec{\sigma} & 0 \\
\end{array}
\right)~~;~~\gamma^5 = \left(
                       \begin{array}{cc}
                         \mathds{I} & 0 \\
                         0 & -\mathds{I} \\
                       \end{array}
                     \right)\,,
                                \ee  in which we find the positive energy spinors
for the mass eigenstates with mass $m_i$ to be
\bea U_i(\vec{k},+1) & = &  \sqrt{\frac{E_i(k)+k}{2k}} \left(
                                             \begin{array}{c}
                                               v(\vec{k},1)  \\
                                               -\frac{m_i}{E_i(k)+k}\, v(\vec{k},1)\\
                                             \end{array}
                                           \right)\,,\label{h1} \\                                            U_i(\vec{k},-1) & = &  \sqrt{\frac{E_i(k)+k}{2k}} \left(
                                             \begin{array}{c}
                                              -\frac{m_i}{E_i(k)+k}\, v(\vec{k},-1)  \\
                                                v(\vec{k},-1)\\
                                             \end{array}
                                           \right)\,,\label{h2}\eea  where $E_i(k)=\sqrt{k^2+m^2_i}$ and the two
component spinors are eigenstates of helicity
\be \vec{\sigma}\cdot \widehat{\vec{k}} \, v(\vec{k},h)= h \, v(\vec{k},h)~;~h=\pm1\,.\ee
The spinors $U$ are normalized to unity and become right handed ($h=1$) or left handed ($h=-1$) respectively
in the ultrarelativistic limit. The propagator in real time is given by
\be   S_i(\vec{k},t) = \int \frac{dp_0}{2\pi } e^{-ip_0 t} S_i(p)\,,\label{RTS}\ee
and for free field theory \be S_i(p) = \frac{\pslash + m_i}{p^2-m^2_i+i0^+}\,. \label{FFS}\ee
It is    straightforward   to confirm that for $t>0$ the result of Eq. (\ref{propover}) is $e^{-iE_i(k)t}$.

In generalizing the above discussion to the case of active-unsterile
mixing, we face two caveats:
\begin{enumerate}
\item the mixing angle $\Phi$ in Eq.
(\ref{mixangles}) depends on the four momentum; this is one of the
features that prompted the study of unsterile-active mixing as a
potential reconciliation of MiniBoone and LSND data as suggested in
Ref. \cite{Schwetz:2007cd},
\item the pole of the unsterile-like eigenstate $\nu_3$ is complex with the imaginary
part being associated with the decay of the sterile neutrino into an active one and quanta
of the hidden conformal sector \cite{Boyanovsky:2009ke}.
\end{enumerate}

A full quantum field theory study along the lines   presented in Refs. \cite{boyaho,wuboya}
 incorporates the full propagators and spectral densities, which are dominated by the poles below
the continuum threshold in the case of the active-like modes
and the complex pole above threshold for the unsterile like mode.
The (four) momentum dependence of the mixing angles is then
evaluated at the position of these poles (see
Refs. \cite{boyaho,wuboya} for a detailed discussion). In these
references, the dynamics of neutrino oscillations was studied as an
initial value problem for an initial wave
packet. However, the real oscillation experiment deals with neutrinos produced at a source and detected
via charged leptons in a far detector, so that the neutrino is virtual in this process.  In this case, the real time quantum field theoretic analysis of oscillation dynamics is more complex. The interaction vertices are
written in terms of the fields that create and annihilate the mass
eigenstates and ``flavor'' neutrinos are an intermediate state,
described by a propagator associated with the mass eigenstates.
Such study is just beginning to emerge, and progress will be reported elsewhere \cite{future}.  A full quantum field theoretical study of active-unsterile dynamics
is beyond the scope of this article, but we can extrapolate from the study in Refs. \cite{boyaho,wuboya}
and  take the mixing angles to be associated with the single (quasi) particle states, therefore
evaluated on the mass shell of the mass eigenstates corresponding to the poles in the respective propagators.

In the case of the unsterile mode, the
imaginary part   of the pole (decay width) is subleading in the
small ratio $\Delta$ (see Eq. (\ref{gama})) and we neglect it in
the evaluation of the mixing angle $\Phi$ for the mass eigenstate
$\nu_3$.

Inverting the relation (\ref{0angle}-\ref{heavyangle}) and
evaluating the mixing angles on the single particle mass shells we
obtain, \bea |\nu_{a1}\rangle & = &  \cos \Phi_2 \cos \theta
|\nu_{2}\rangle + \sin \Phi_3 \cos \theta |\nu_{3}\rangle -\sin
\theta |\nu_{1}\rangle\,, \label{active1} \\|\nu_{a2}\rangle & = & \cos
\Phi_2 \sin \theta |\nu_{2}\rangle + \sin \Phi_3 \sin \theta
|\nu_{3}\rangle +\cos \theta
|\nu_{1}\rangle\,, \label{active2} \\
|\nu_{U}\rangle & = &  -\sin \Phi_2  |\nu_{2}\rangle + \cos \Phi_3  |\nu_{3}\rangle \,,  \label{unste}\eea where
to leading order in $\Delta$ we find \be  \sin \Phi_2   \simeq \Bigg[ \frac{\Delta}{2[1+\frac{\Delta}{2} ]} \Bigg]^{\frac{1}{2}} ~~;~~
\cos \Phi_2   \simeq   \Bigg[ \frac{1}{  1+\frac{\Delta }{2} }
\Bigg]^{\frac{1}{2}}\,,\label{coseno2}\ee \be \sin \Phi_3 =
\frac{\delta}{\sqrt{1+\delta^2}}~~;~~ \cos \Phi_3 =
\frac{1}{\sqrt{1+\delta^2}}\,,\label{coseno3}\ee and  \be \delta =
\sqrt{\frac{\Delta}{2}}\,\Bigg[ \Delta^\frac{1}{1-\eta}\, \cos\Big(
\frac{\pi \eta}{1-\eta}\Big)\Big] \Bigg]^\frac{\eta}{2}\,.
\label{smaldelta}\ee We will focus on negative helicity positive
energy states, for which the spinor wave functions are given by Eq.
(\ref{h2}).

An immediate caveat of this formulation is that as a consequence of
the four-momentum dependence of the mixing angle $\Phi$, the states
$\nu_{a1,a2,U}$ introduced above as a straightforward
generalization of the familiar quantum mechanical description, are
\emph{not orthogonal}, despite the orthogonality of the mass
eigenstates $\nu_{1,2,3}$. This is because $\Phi_2\neq \Phi_3$. This is an
unavoidable consequence of the energy-momentum dependence of the
mixing angle and of the effort to establish a correspondence with the
familiar single particle quantum mechanical description of an
inherently many particle problem. The non-orthogonality of these
states is \emph{small} for $\Delta \ll 1$ as manifest by Eqs.
(\ref{coseno2},\ref{coseno3},\ref{smaldelta}).

While the  single particle quantum
mechanical analogy has limitations in absence of a full quantum field theory
treatment of neutrino mixing and oscillations directly \emph{in real time}, we adopt the
approximate single particle description afforded by (\ref{active1},\ref{active2},\ref{unste}) as a \emph{proxy} description and
proceed to explore this formulation as a prelude towards a firmer
understanding. A thorough field theoretical description of real time oscillations is postponed for future work.

The fields associated with the mass eigenstates $\psi_i, i=1,2,3$
are expanded at $t=0$ just as in Eq. (\ref{psiqua}) but with the
spinors $U,V$ solutions of the Dirac equations with masses
$$
\mathcal{M}_1=0, ~~\mathcal{M}_2= \frac{1}{2}M {\Delta}   , ~~
\mathcal{M}_3 = M \Big[1 + \Delta^\frac{1}{1-\eta}\, \cos\Big(
\frac{\pi \eta}{1-\eta}\Big)\Big]^\frac{1}{2}\,, $$
respectively \cite{Bjorken:1979dk}.  The annihilation operators
$b(\vec{k},h),\, d(\vec{k},h)$ and creation operators
$b^\dagger(\vec{k},h), \,d^\dagger(\vec{k},h)$ are interpolating
Heisenberg field operators   whose time evolution is determined by the total
Hamiltonian such that \cite{Bjorken:1979dk} \be \langle 0|\,
\psi_i(\vec{k},t) \overline{\psi}_i(\vec{k},0) \,|0\rangle =
S_i(\vec{k},t>0)\,. \label{vacexp}\ee

In order to carry out the integration in $p_0$ as in Eq. (\ref{RTS}), it is convenient to write the propagators
in terms of their spectral representation \cite{Bjorken:1979dk} \be S_i(p) = \int_0^\infty dQ^2 ~\frac{\pslash \,\rho^{(1)}_i(Q^2)+\mathcal{M}_i \,\rho^{(2)}_i(Q^2)}{p^2-Q^2+i0^+}\,, \ee
from which we find \be S_i(\vec{k},t >0) = \int_0^\infty dQ^2 ~ \frac{\widetilde{\pslash} \,\rho^{(1)}_i(Q^2)+
\mathcal{M}_i \,\rho^{(2)}_i(Q^2)}{2E_i(Q)} \, e^{-i\mathcal{E} (Q)t}\,, \label{Softi}\ee where \be \mathcal{E}(Q)=
\sqrt{k^2+Q^2} ~~;~~\widetilde{\pslash} = \gamma_0 \,\mathcal{E}(Q)-\vec{\gamma}\cdot\vec{k} \,.\label{quans}\ee
To leading order in $\Delta$, we have $\rho_i^{(1)}(Q^2) = \rho_i^{(2)}(Q^2)~~;~~i=2,3$. For
the massive active-like mode \be \rho^{(1)}_2(Q^2)= Z_2 \, \delta(Q^2-\mathcal{M}^2_2) + \frac{\rho_2(x)}{M^2}~~;~~
x= \frac{Q^2-M^2}{M^2}\,,\label{rhof2}\ee where $Z_2$ is given by Eq. (\ref{Z1}) and $\rho_2(x)$
by Eq. (\ref{disc1}). For the unsterile like mode \be \rho_3^{(1)}(Q^2) = \frac{\rho_3(x)}{M^2} \,,\label{rhof3}\ee
where $\rho_3(x)$ is given by Eq. (\ref{disc2}). Near the complex pole in the continuum, the propagator for the unsterile-like neutrino
can be approximated the Breit-Wigner form
Eq. (\ref{alfaminbeta}), which for $\Delta, \eta \ll 1$ describes a narrow resonance near the real axis.

For long times, the integral over the dispersive variable is dominated by the pole and the continuum threshold.
The technical details of the evaluation of the threshold contribution are  found in the appendix (see Eq. (\ref{threshcont})).
For negative helicity states, we find
\bea \langle \nu_1 |\, e^{-iHt}\, |\nu_1 \rangle & = & e^{-iE_1(k) t}\,, \label{nu1oft}\\
\langle \nu_2 |\, e^{-iHt}\, |\nu_2 \rangle & = & Z_2\,e^{-iE_2(k) t}+ \frac{A_2}{2\,i^{1+2\eta}}
\Bigg( \frac{E_M}{M^2\,t}\Bigg)^{1+2\eta} \,e^{-iE_M(k)t}\,, \label{nu2oft}\\
\langle \nu_2 |\, e^{-iHt}\, |\nu_2 \rangle & = &
Z_3\,e^{-iE_3(k)t}\,e^{-\frac{\Gamma\,t}{2\gamma} }\,,
\label{nu3oft}\eea where $E_i(k)=\sqrt{k^2+\mathcal{M}^2_i}$, $\Gamma$ is given by Eq. (\ref{gama}),
$\gamma = E_3(k)/\mathcal{M}_3$ is the Lorentz time dilation factor
and (see appendix, Eq. (\ref{A2coef})) \be
A_2 = \frac{\Delta^2}{4 \pi} \, \sin (2 \pi \eta)  \, \Gamma(1+2\eta) \,. \label{A2co}
\ee The extra factor $1/2$ in
(\ref{nu2oft}) as compared to the bosonic case studied in the
appendix (see Eq. (\ref{threshcont})) is a result of the spinor
overlaps in the limit $\Delta \ll 1$.

\begin{figure}
\begin{center}
\includegraphics[width=10cm,keepaspectratio=true]{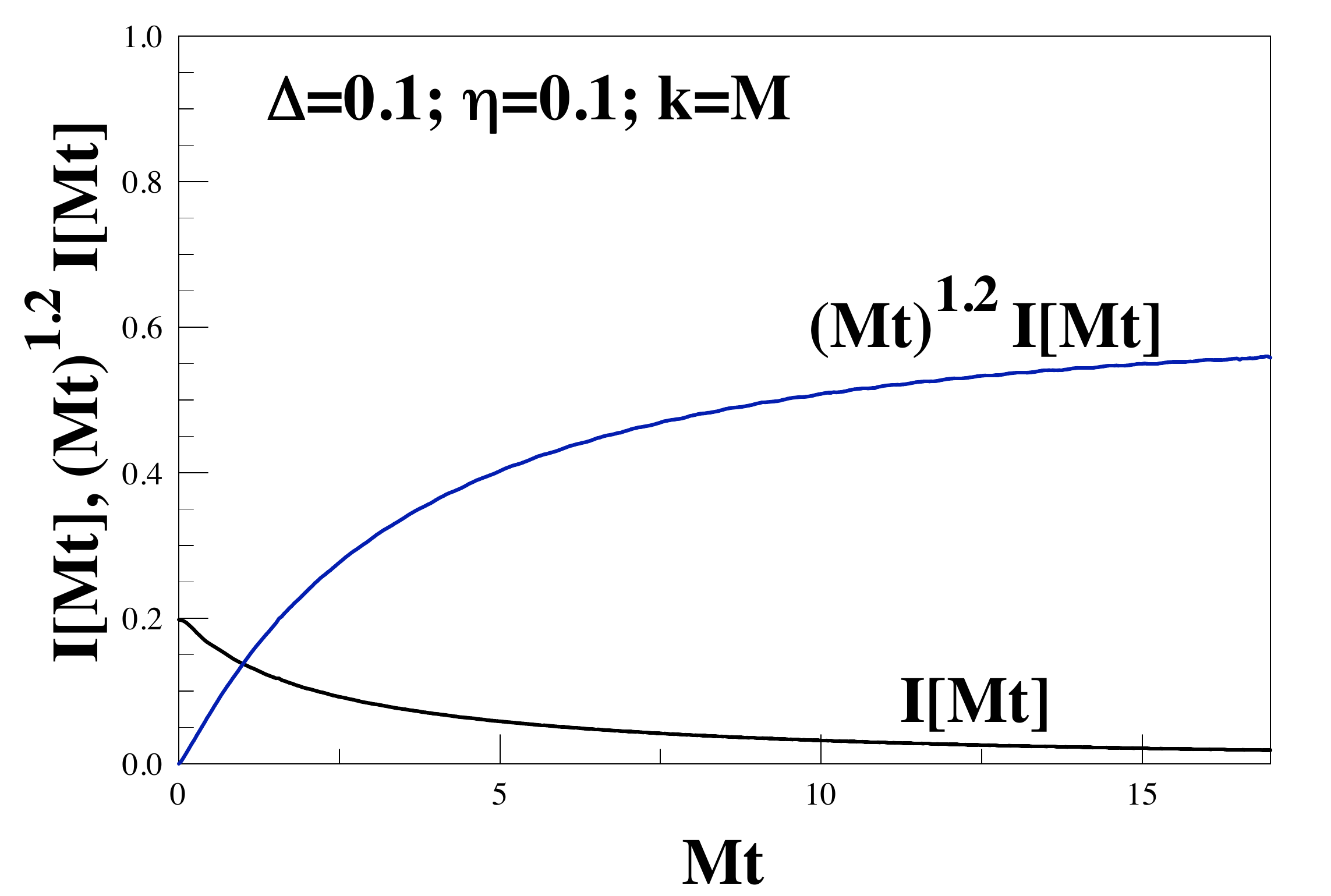}
\caption{The continuum integral $I[Mt]= |\langle
\nu_2|\,e^{-iHt}\,|\nu_2\rangle |_{cont}/A_2$.} \label{rho2time}
\end{center}
\end{figure}

The result in Eq. (\ref{nu2oft}) is valid in the long time limit when the threshold
contribution dominates the integral over the continuum \cite{brown}. Fig. (\ref{rho2time}) displays the integral over the continuum and confirms its asymptotic long time
limit.

We are now in position to obtain the active neutrino disappearance
and appearance probabilities of active-like neutrinos.

\subsection{Disappearance probabilities} The disappearance transition
amplitudes are given by \bea  \langle  \nu_{a1}
|\,e^{-iHt}\,|\nu_{a1} \rangle  &=& \sin^2\theta \langle \nu_{
1}\,|e^{-iHt}\,|\nu_{1} \rangle + \cos^2\Phi_2 \cos^2\theta
\,\langle \nu_{2}\,|e^{-iHt}\,|\nu_{2}
\rangle \nonumber \\
& &+\, \sin^2\Phi_3 \cos^2 \theta \, \langle
\nu_{3}\,|e^{-iHt}\,|\nu_{3} \rangle\,,  \label{nua1amp}\\
\langle
\nu_{a2}\,|e^{-iHt}\,|\nu_{a2} \rangle  &=& \cos^2\theta \langle \nu_{ 1}\,|e^{-iHt}\,|\nu_{1} \rangle +  \cos^2\Phi_2 \sin^2\theta
\,\langle \nu_{2}\,|e^{-iHt}\,|\nu_{2} \rangle \nonumber \\
& & +\, \sin^2\Phi_3 \sin^2
\theta \, \langle \nu_{3}\,|e^{-iHt}\,|\nu_{3} \rangle \,,
\label{nua2amp}\eea where the transition amplitudes $\langle \nu_{i
}\,|e^{-iHt}\,|\nu_{i} \rangle\,;i=1,2,3$ are given by Eqs.(\ref{nu1oft}-\ref{nu3oft}). In obtaining the probabilities there are several interference terms which manifest themselves on different time scales.
We will only keep the \emph{slow} oscillatory terms involving the differences $$E_1(k)-E_2(k) \approx \mathcal{M}^2_2 / 2 k \qquad {\rm and}
\qquad E_M(k)-E_3(k) \approx (\mathcal{M}^2_3 -M^2)/2E_M(k)$$ in the ultrarelativistic limit, and neglect phases that oscillate on much
shorter time scales, since these average out. As is customary, we replace time by the baseline $t \rightarrow L$, and define
\be \tilde{A}(L) = \frac{A_2}{2} \Bigg[
\frac{E_M(k)}{M^2L}\Bigg]^{1+2\eta} ~~;~~\xi =
\frac{\pi}{2}(1+2\eta) \,. \label{atilde}\ee We then have \bea
\mathcal{P}_{a_1 \rightarrow a_1}(L) &=&  Z^2_2\cos^4\Phi_2
\cos^4\theta + \sin^4\theta+ \frac{Z_2}{2} \cos^2\Phi_2
\sin^22\theta \cos \left[ \frac{\mathcal{M}^2_2\,L}{2k}\right] +\, \tilde{A}^2(L) \cos^4\Phi_2\cos^4\theta \nonumber \\
&&
+\, 2\, \tilde{A}(L) \, Z_3 \cos^2\Phi_2 \sin^2\Phi_3 \cos^4\theta \,e^{-\frac{\Gamma}{2\gamma} L}\,
\cos\left[ \frac{(M^2-\mathcal{M}^2_3)L}{2E_M(k)}+ \xi\right] \nonumber \\
& & + \, Z^2_3\sin^4\Phi_3 \cos^4\theta \, e^{-\frac{\Gamma}{\gamma} L} \,, \label{pa1a1}
\eea
\bea
\mathcal{P}_{a_2 \rightarrow a_2}(L) &=&  Z^2_2\cos^4\Phi_2 \sin^4\theta + \cos^4\theta+ \frac{Z_2}{2} \cos^2\Phi_2 \sin^22\theta
\cos\left[ \frac{\mathcal{M}^2_2\,L}{2k}\right] +\, \tilde{A}^2(L) \cos^4\Phi_2\sin^4\theta \nonumber \\
&&
+\, 2\, \tilde{A}(L) \, Z_3 \cos^2\Phi_2 \sin^2\Phi_3 \sin^4\theta \,e^{-\frac{\Gamma}{2\gamma} L}\,
\cos\left[ \frac{(M^2-\mathcal{M}^2_3)L}{2E_M(k)}+ \xi\right]  \nonumber \\
& &  + \, Z^2_3\sin^4\Phi_3 \sin^4\theta \, e^{-\frac{\Gamma}{\gamma} L} \label{pa2a2}\,.
\eea

\subsection{Appearance probability}
From the transition amplitude \be  \langle  \nu_{a1}\,|e^{-iHt}\,|\nu_{a2} \rangle  = \frac{1}{2}
\sin 2\theta   \Big\{
\cos^2\Phi_2   \,\langle \nu_{2}\,|e^{-iHt}\,|\nu_{2}
\rangle + \sin^2\Phi_3   \, \langle
\nu_{3}\,|e^{-iHt}\,|\nu_{3} \rangle - \langle \nu_{
1}\,|e^{-iHt}\,|\nu_{1} \rangle \Big\}\,, \label{nua1a2amp}\ee by keeping only the
\emph{slow} interference terms and again replacing $t\rightarrow L$, where $L$ is the baseline, we find
\bea \mathcal{P}_{a_1 \rightarrow a_2}(L) = \mathcal{P}_{a_2 \rightarrow a_1}(L) &=&  \frac{\sin^22\theta}{4} \Bigg\{1 +   \left[Z^2_2+\tilde{A}^2(L) \right] \cos^4\Phi_2  - 2 Z_2 \cos^2\Phi_2 \cos\left[ \frac{\mathcal{M}^2_2\,L}{2k}\right] \nonumber \\
&&  + \, 2 \, \tilde{A}(L) \,Z_3 \cos^2 \Phi_2 \sin^2\Phi_3 \,e^{-\frac{\Gamma}{2\gamma} L}\,\cos\Big[ \frac{(M^2-\mathcal{M}^2_3)L}{2E_M(k)}+ \xi\Big]
 \nonumber \\&&
+\, Z^2_3\sin^4\Phi_3  \, e^{-\frac{\Gamma}{\gamma} L} \Bigg\} \,,\label{pa1a2}\eea

where for $\Delta \ll 1$, it follows that \be  \mathcal{M}^2_3-M^2 =   M^2 \Delta^\frac{1}{1-\eta}\, \cos\Big( \frac{\pi \eta}{1-\eta}\Big)\,. \ee

\subsection{Consequences of the anomalous dimension} The unsterile neutrino is characterized by
the anomalous dimension $\eta$, which is responsible for $\tilde{A},Z_2,Z_3,\Gamma \neq 0$ \emph{and}
 $\Phi_3\neq \Phi_2$. Therefore it is important to quantify the potential phenomenological effects of the
 appearance and disappearance probabilities associated with a non-vanishing (and perhaps large) anomalous
 dimension.  For $\eta =0$, these transition probabilites are the following (we neglect terms that oscillate on the
 time scale $1/M\ll k/\mathcal{M}^2_2$ and average out)
\be  \mathcal{P}_{a_1\rightarrow a_1} = \big(\cos^2 \Phi \cos^2
\theta + \sin^2\theta \big)^2+ \sin^4\Phi \cos^4 \theta - \cos^2\Phi
\sin^2(2\theta) \sin^2\Big[\frac{\mathcal{M}^2_2L}{4k} \Big]\,,
\label{pa1pa1eta0}\ee \be  \mathcal{P}_{a_2\rightarrow a_2} =
\big(\cos^2 \Phi \sin^2 \theta + \cos^2\theta \big)^2+ \sin^4\Phi
\sin^4 \theta - \cos^2\Phi \sin^2 2\theta
\sin^2\Big[\frac{\mathcal{M}^2_2L}{4k} \Big]\,,
\label{pa2pa2eta0}\ee \be \mathcal{P}_{a_1\rightarrow a_2}
=\mathcal{P}_{a_2\rightarrow a_1} = {\sin^2 2\theta} ~\Bigg\{
\frac{\sin^4\Phi}{2}+
\cos^2\Phi \sin^2\Big[\frac{\mathcal{M}^2_2L}{4k} \Big]  \Bigg\}\, , \label{pa1a2eta0} \ee where \be \sin\Phi = \frac{m}{\sqrt{M^2+m^2}} \,.
 \label{sinFi}\ee

 In the above expressions, we have neglected interference terms that
 oscillate on the fast time scale $\propto 1/M$ which average out
 on   the longer time scales of the oscillatory contributions
 displayed in these expressions\footnote{This is the reason that $\mathcal{P}_{a_1 \rightarrow a_2}$ does not vanish at $t=0$.}.

We note that $ \mathcal{P}_{a_1\rightarrow a_1}\neq
\mathcal{P}_{a_2\rightarrow a_2}$. This is a consequence of the fact
that the transformation between flavor and mass eigenstates is
\emph{not unitary} in the active sector. In principle the existence
of a \emph{canonical} ($\eta=0$) sterile neutrino may be
experimentally determined by measuring the disappearance probability
for \emph{both} active species (by measuring the associated charged
leptons) along with the appearance probability. The difference in the
disappearance probability for both active species signals the presence of a
``sterile'' degree of freedom that enters in the definition of the
mass eigenstates but not in the weak interaction vertices. The
combined measurement of all three probabilities for fixed baseline
and energy would allow us to extract both mixing angles and oscillation
lengths.

\begin{figure}
\begin{center}
\includegraphics[width=8cm,keepaspectratio=true]{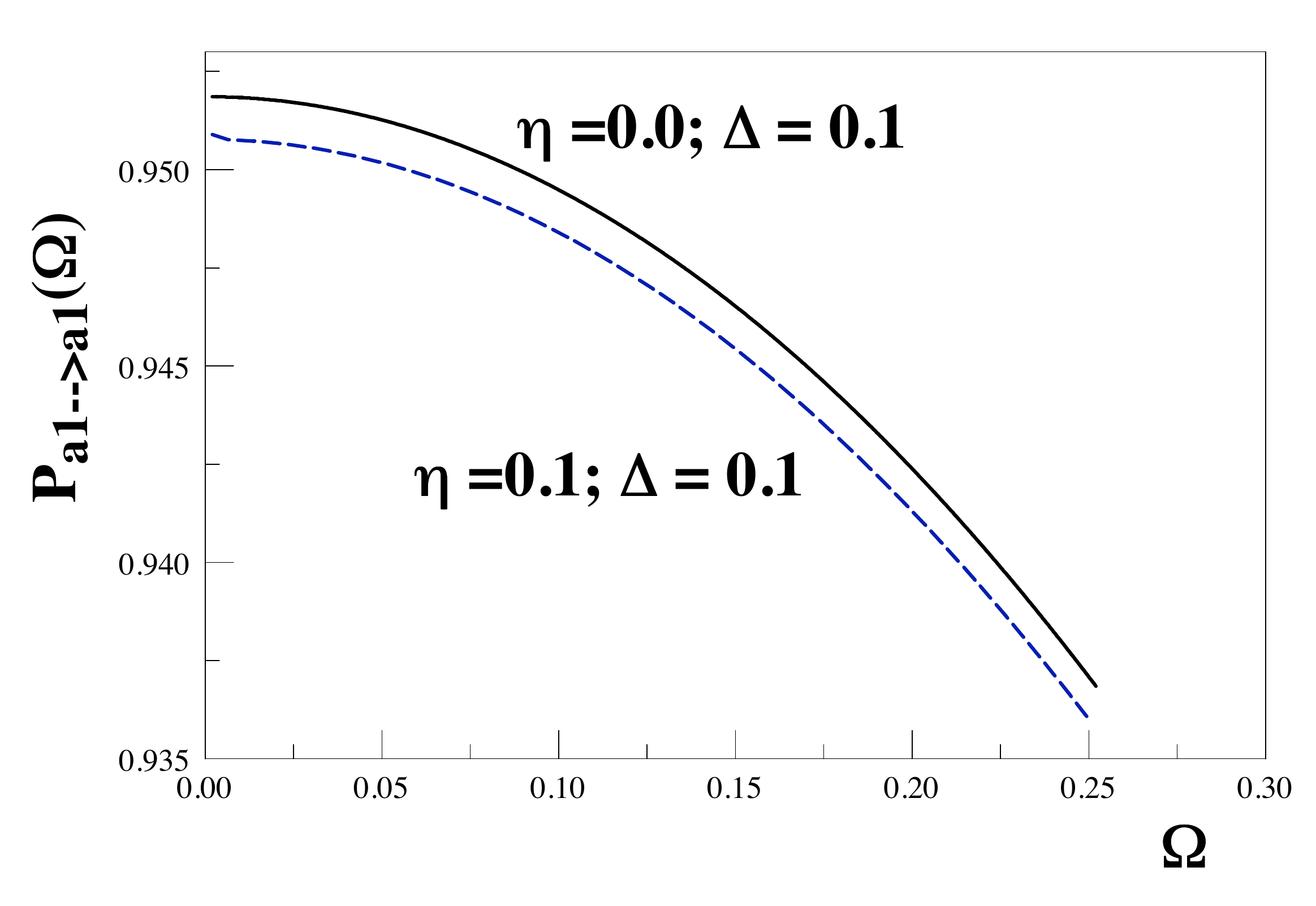}
\includegraphics[width=8cm,keepaspectratio=true]{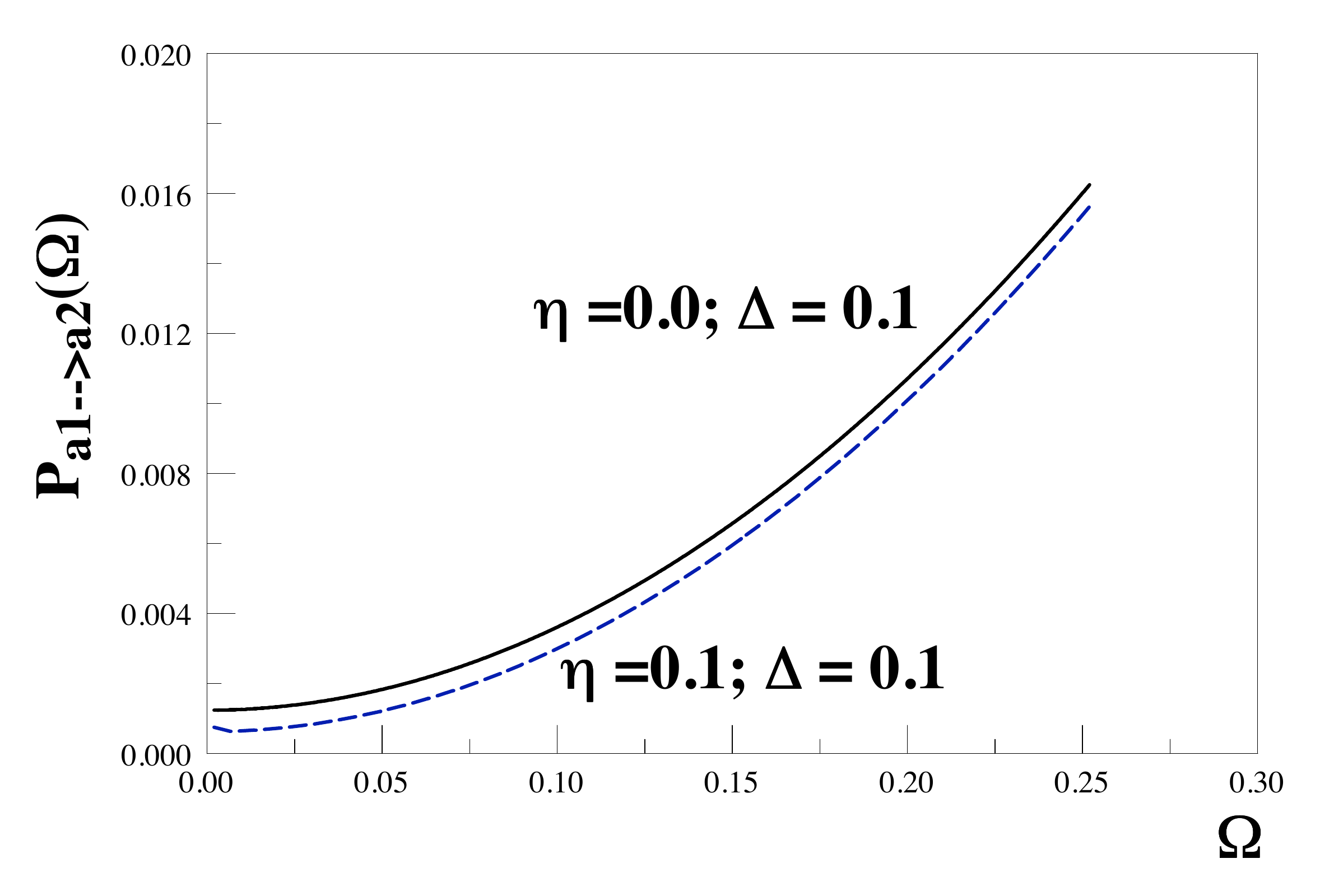}
\caption{Disappearance and appearance probabilities for $\Delta
=0.1$ and $\eta =0,0.1$ to compare canonical and unsterile
neutrinos.} \label{anomalous}
\end{center}
\end{figure}

\begin{figure}
\begin{center}
\includegraphics[width=8cm,keepaspectratio=true]{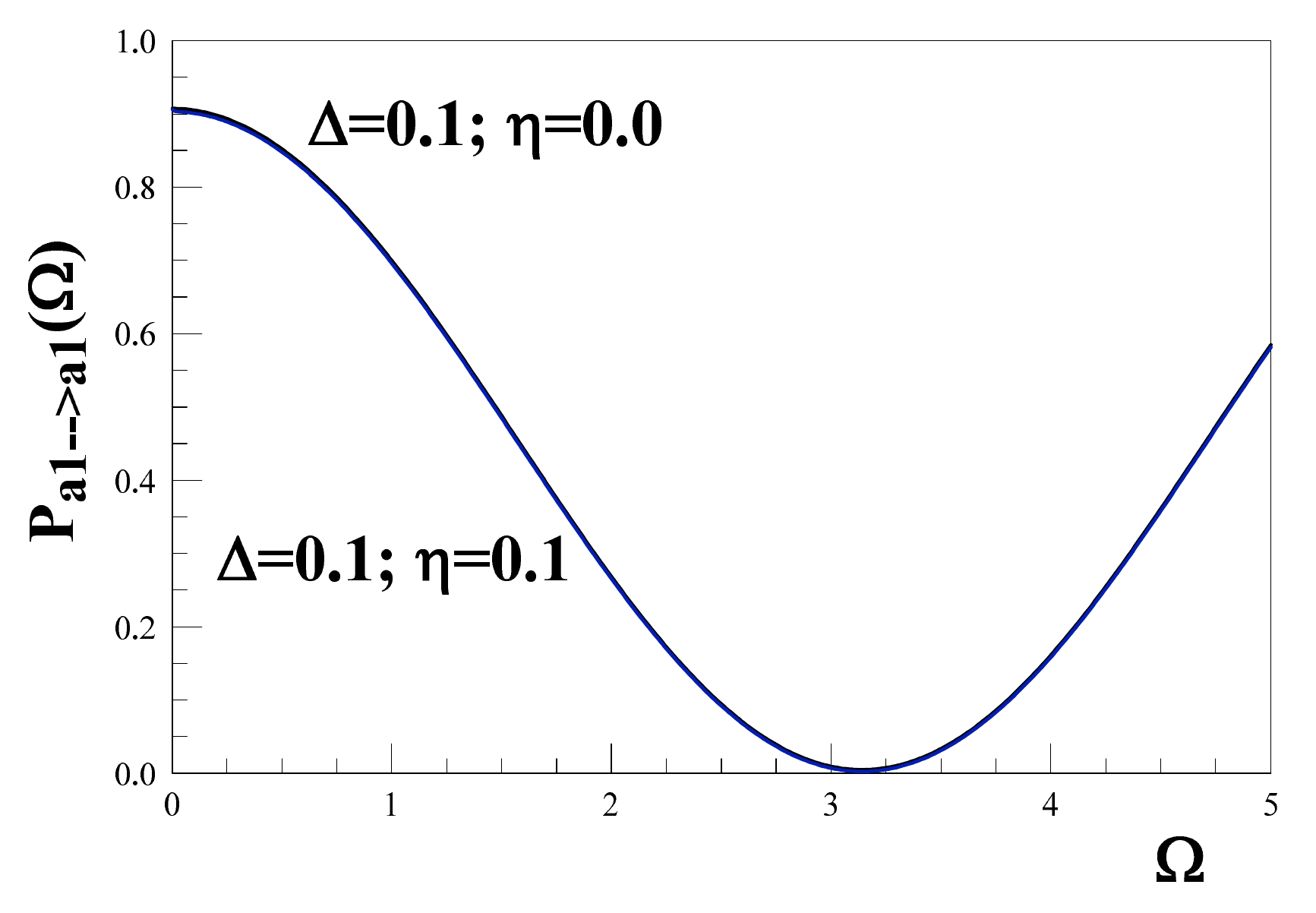}
\includegraphics[width=8cm,keepaspectratio=true]{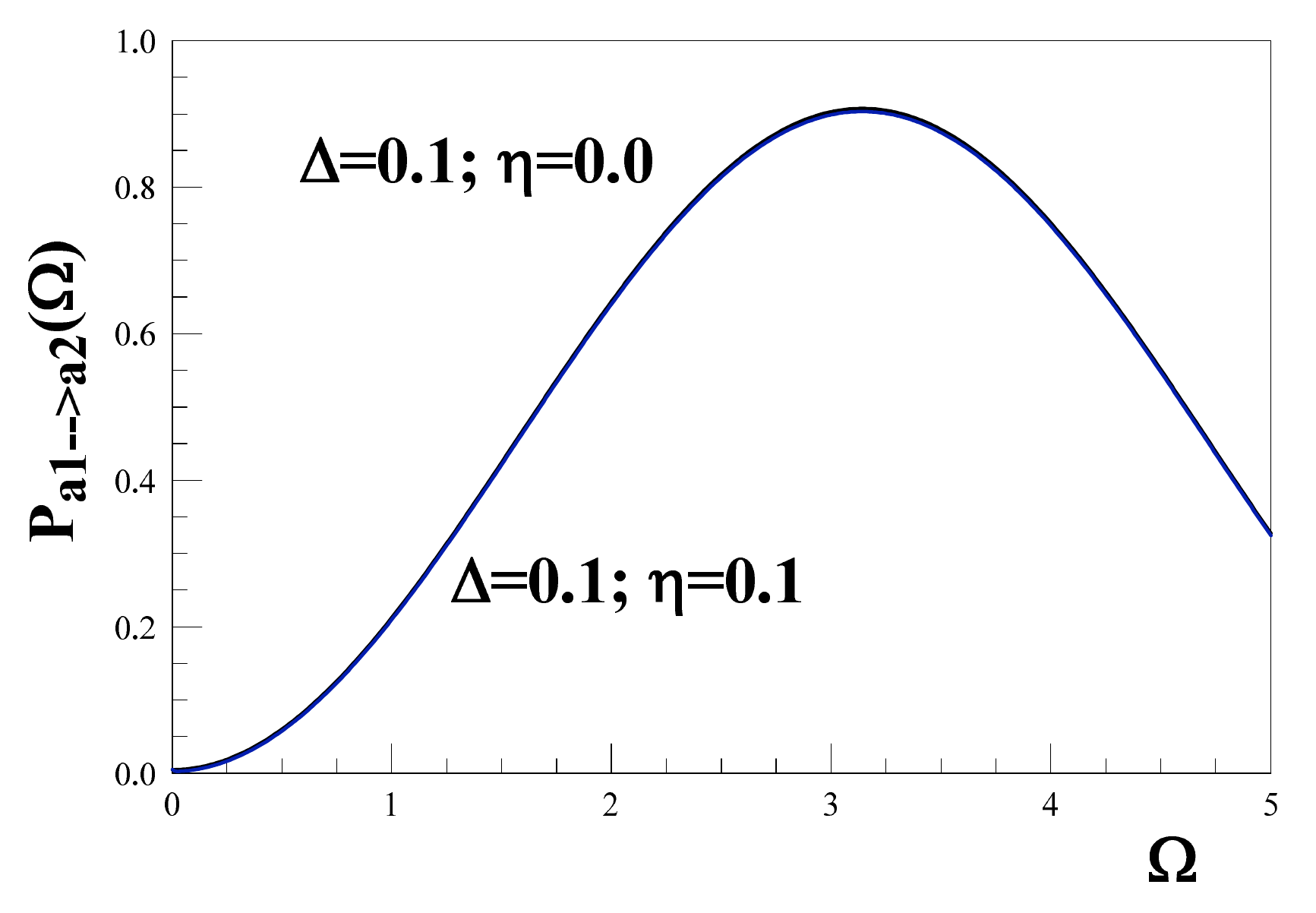}
\caption{By the first oscillation peak, the difference between canonical sterile and unsterile neutrinos has disappeared.} \label{firstpeak}
\end{center}
\end{figure}

A non-vanishing anomalous dimension introduces different angles
$\Phi_{2,3}$ as a consequence of the momentum dependence of the
mixing angle. This then gives rise to new oscillatory contribution with a different
oscillation length that is also multiplied by an attenuation factor.
To understand the different time scales it is convenient to
introduce \be \Omega = \frac{\mathcal{M}^2_2 L}{2k} = \frac{M^2
\Delta^2 L}{8k} \equiv 2\pi \frac{L}{L_{osc}}\,, \label{omega}\ee where
$L_{osc}$ is the usual active oscillation length, as well as $\widetilde{\Omega}$,  the new
oscillation frequency arising from the interference between the
massive active-like threshold and the unsterile-like pole \be
\widetilde{\Omega} = \frac{(\mathcal{M}^2_3 -M^2)L}{2k} = \Omega
\Bigg[\frac{4}{\Delta^{\frac{1-2\eta}{1-\eta}}}\,\cos\Big[ \frac{\pi
\eta}{1-\eta}\Big] \Bigg]\,, \label{tildeomega}\ee in terms of which we
find \be \frac{\Gamma L}{2\gamma} = \widetilde{\Omega}~ \tan\Big[
\frac{\pi \eta}{1-\eta}\Big]\,. \label{atetil}\ee
For $\Delta \ll 1$,
the oscillation frequency $\widetilde{\Omega}$  is \emph{larger}
than the active oscillation frequency $\Omega$. As a consequence of
this discrepancy in the frequency of the oscillatory
contributions, as well as the attenuation and the wave-function
renormalization factors,  it follows that both the
disappearance and appearance probabilities are \emph{suppressed}
relative to the case of a canonical sterile neutrino. However,
because of the attenuation factors, the suppression is substantial
only  for \emph{short baseline experiments}, namely $\Omega \ll 1$.
This suppression is displayed in Fig. (\ref{anomalous}). We can also see in Fig. (\ref{firstpeak}) that by the first oscillation peak, the novel oscillatory contributions are well suppressed that it will be difficult to differentiate between canonical sterile and unsterile neutrinos.

The discussion above applies in the limit $\Delta \ll 1$ which has been invoked from the
beginning to establish a see-saw hierarchy between the active-like and unsterile-like neutrinos. However, the proposal for a ``solution'' of the LSND/MiniBooNE discrepancy introduces a  sterile neutrino
in the $eV$ mass range, namely within the same mass range as the active neutrinos. This possibility requires that $\Delta \sim \mathcal{O}(1)$.
Although our study does not apply directly to this regime, we can extrapolate to this regime in several aspects.

First, the mixing angle $\Phi_3$, which determines the overlap between active-like and unsterile-like modes,
becomes of the \emph{same} order as the angle $\theta$ that determines the overlap between active-like states. This modifies
the appearance and disappearance probabilities by overall normalizations. However, the argument on Lorentz invariance still
implies, at least within the quantum mechanical description of oscillations, that the angles are evaluated on the mass shell of the single particle states and \emph{do not} depend on the energy.

\begin{figure}
\begin{center}
\includegraphics[width=8cm,keepaspectratio=true]{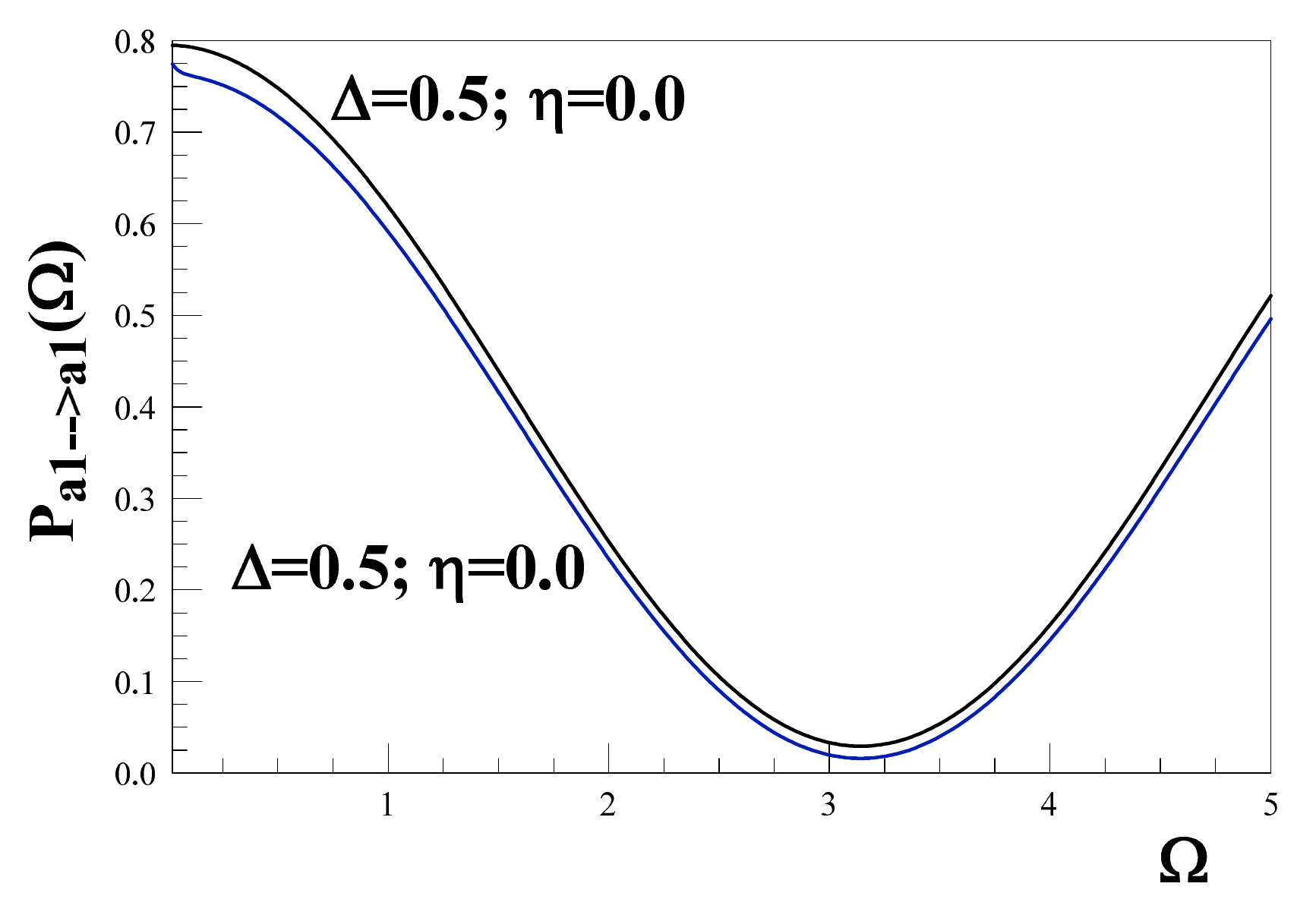}
\includegraphics[width=8cm,keepaspectratio=true]{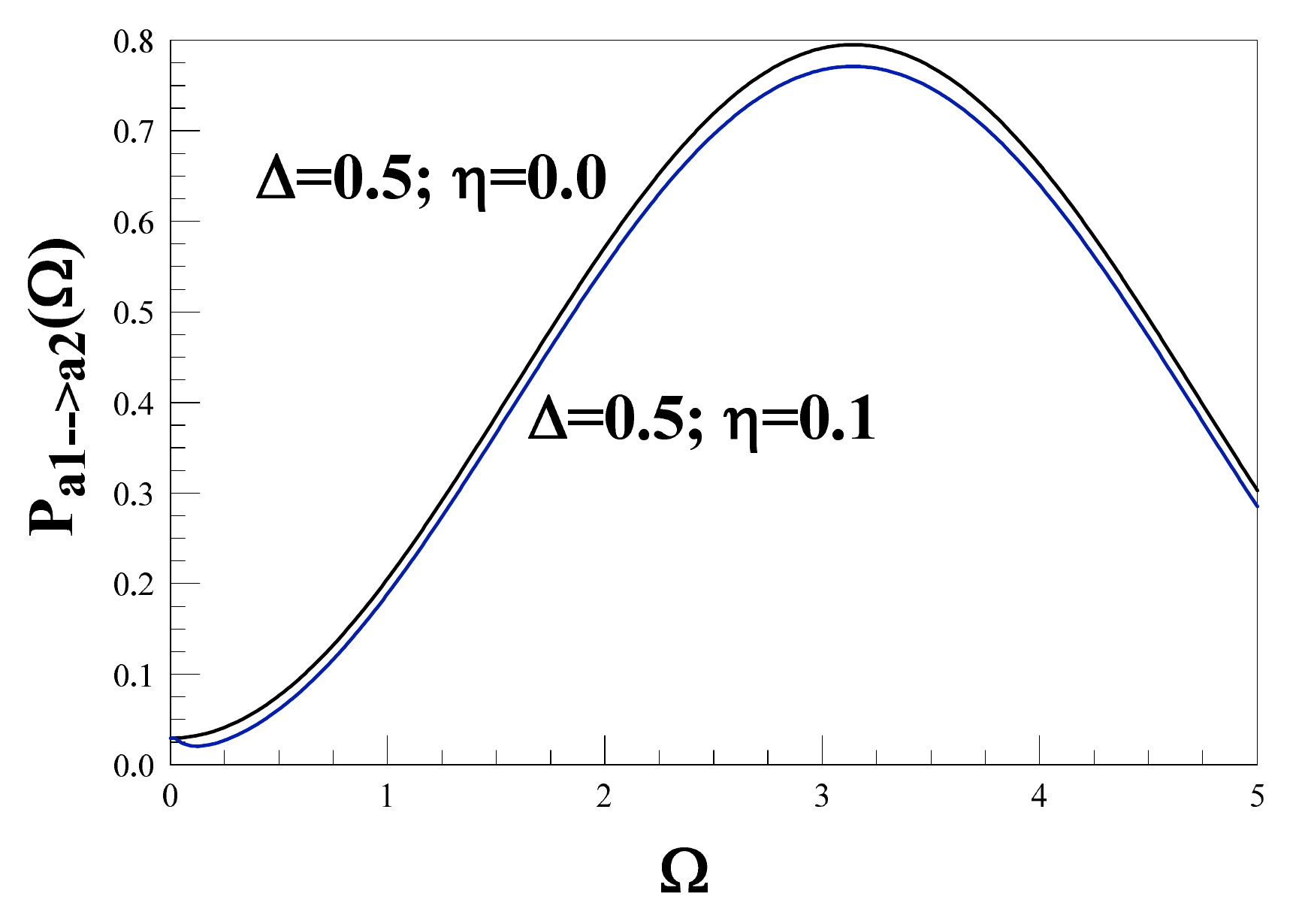}
\caption{For larger $\Delta$ the suppression effect due to the anomalous dimension is more prominent.} \label{largedelta}
\end{center}
\end{figure}

For $\Delta \sim \mathcal{O}(1)$, it follows from Eq. (\ref{tildeomega}) that $\tilde{\Omega} \sim \Omega$ and the oscillation frequencies of both oscillatory terms in the probabilites are of the same order. Furthermore, the factor $\tilde{A}
$ of Eq. (\ref{atilde}) that determines the interference
terms from the threshold of the massive active mode and the unsterile mode become of order $\sin[2\pi\eta] $. Therefore, for
 $\eta \lesssim 1/3$, but not too small, this overlap yields a potentially interesting energy dependence. In this case, the suppression of the probabilities as a consequence of the anomalous dimension is non-vanishing even for long baseline events, namely with $\Omega \sim \mathcal{O}(1)$.  However, for
  large  $\eta$ such that $\tan \pi \eta \sim 1$  and $\Delta \sim \mathcal{O}(1)$, the attenuation length of the overlap (see Eq. (\ref{atetil}))
   is substantial and the interference between active and unsterile-like is suppressed over the baseline. The appearance and disappearance probabilities for $\Delta =0.5$ are shown in Fig. (\ref{largedelta}).
   
\section{Summary}

In this article, we considered the possibility that the $SU(2)$ singlet sterile neutrino is an unparticle and mixes with
 the active neutrinos. The unparticle sterile neutrino -- or unsterile neutrino -- is an interpolating field that describes a multiparticle
  continuum as a consequence of coupling to a ``hidden'' conformal sector and whose correlation functions feature an anomalous scaling dimension
  $\eta$. We analyzed  the consequences of its mixing with two active neutrinos via a ``minimal" see-saw mass matrix, with massless
  active neutrinos and no active-active mixing term.

We have introduced a generalization of the usual
quantum mechanical description of the dynamics of mixing and oscillation that incorporates the propagators for the fields associated
 with the mass eigenstates and includes off-shell corrections to the time dependence of the appearance and disappearance probabilities
 for the active-like neutrinos.
 We find a remarkable interference phenomenon between the massive active-like and unsterile-like modes as a consequence of threshold
 effects which modify both appearance and disappearance probabilities in a novel manner.
 
 The presence of the anomalous dimension that gives the unparticle nature to the sterile neutrino has profound consequences in the
 disappearance and appearance probabilities for the active-like modes. For a canonical sterile neutrino, the disappearance probability
 is different for the different flavors as a consequence of the fact that the transformation between flavor and mass eigenstates is not
  unitary in the active sector. This same feature remains in case of the unsterile neutrino, however, the anomalous dimension is responsible
   for novel time dependent phenomena corresponding to the new oscillatory term arising from the interference of the threshold of the massive
    active-like mode and the unsterile mode. This oscillatory term is multiplied by an attenuating function of baseline but the oscillation
    length is different from the active-active one. These novel contributions are consequences of the unparticle nature of the
    sterile neutrino and result in a \emph{suppression} of both probabilities on short baseline experiments as compared to the
     case of active-sterile mixing with a canonical sterile neutrino. Combined analysis of short and long baseline experiments may, therefore,
     provide a  diagnostic tool for the unparticle nature of a sterile neutrino.  Another important manifestation of the unparticle
     nature of the sterile neutrino is the  non-vanishing spectral density ``inherited'' by the massive active mode, which
     may lead to the possibility of new reaction channels when the standard model interactions are present.

Although these effects would be potentially relevant in the
phenomenological reconciliation between the LSND and MiniBooNE data,
we find that as a consequence of Lorentz invariance, the mixing
angles, when evaluated on the mass shell of the corresponding mass
eigenstates \emph{do not } depend on energy as required for a
reconciliation of the data\cite{Schwetz:2007cd}. Furthermore, within
the \emph{simple model} considered here, we find that either with a
large see-saw hierarchy of masses between the unsterile-like and the
active like neutrinos, or even if they all feature the same mass
scale, the new interference terms that could be responsible for
novel effects are probably too small to reconcile
 the LSND and MiniBooNE data.

We would like to emphasize that there are, however, some important caveats: the framework introduced as a generalization of the
quantum mechanical description of the dynamics of mixing and oscillations can at best be a \emph{proxy} for a full quantum field
theoretical treatment of the appearance and disappearance probabilities in real time in which ``flavor'' neutrinos are intermediate states,
including the propagation of the unsterile degree of freedom. Furthermore, considering either a more realistic $3 +\tilde{1}$ scenario or a
more general mixing matrix in $2+\tilde{1}$ scheme with mass scales of the same order may provide a  scenario in which the novel phenomena
found here may yield rich phenomenology. For example, in $2+\tilde{1}$ scheme, giving one of the active neutrinos a mass, which can be acquired
 by including another type beyond standard model sector,  already results in a momentum dependent active-active mixing angle.

Although the simple model studied here may not provide the reconciliation between the LSND and MiniBooNE data, the novel time dependent phenomena
that emerges as a consequence of the unparticle nature of the sterile neutrino warrants further and deeper study. We anticipate important
cosmological consequences in the equilibration of active neutrinos, novel mechanisms for sterile neutrino production and possibly
interesting consequences for light dark matter candidates.

\begin{acknowledgments}
D.B. acknowledges support from the U.S. National Science Foundation through Grant No.
PHY-0553418 and PHY-0852497. R. H. and J. H. are supported by the DOE through Grant No. DE-FG03-91-ER40682.
The authors thank T. Schwetz for correspondence and interesting suggestions that prompted this study.
\end{acknowledgments}

\appendix

\section{Oscillations of bosonic single particle states}

Neutrino mixing and oscillations is conventionally studied via a single particle quantum mechanical description. In this formulation, there is no distinction on whether the states of the quantum field theory correspond to fermionic or bosonic particles, and in the case of fermions,
there is no mention of helicity or any other quantum numbers besides energy and momentum.

In this appendix, by considering bosonic particles, we study the dynamics of oscillations to establish potential differences with the fermionic case described in the text.  Let us consider a mixing of two canonical neutrinos. The mass eigenstates are given by
\bea
\nu_1 & = &  \cos \theta \, \nu_a + \sin \theta \, \nu_b, \label{psi1}\\
\nu_2 & = &  \cos \theta\, \nu_b - \sin \theta\, \nu_a\,, \label{psi2}
\eea
with $\nu_{a,b}$ are the flavor eigenstates and the masses corresponding to the fields $\nu_{1,2}$ are $M_{1,2}$, respectively. The transition amplitude $|\langle \nu_a(t)|\nu_b(0) \rangle|^2$ is calculated using ordinary quantum mechanics:
\be
\langle \nu_a(t)|\nu_b(0) \rangle = \langle \nu_a(t)|\nu_1(t) \rangle \, \langle \nu_1(t)|\nu_1(0) \rangle \, \langle \nu_1(0)|\nu_b(0) \rangle +  \langle \nu_a(t)|\nu_2(t) \rangle \, \langle \nu_2(t)|\nu_2(0) \rangle \, \langle \nu_2(0)|\nu_b(0) \rangle,
\ee
where the equal time overlaps $\langle \nu_{a,b}|\nu_{1,2} \rangle$ are read from Eqs. (\ref{psi1}) and (\ref{psi2}), while the time evolution $\langle \nu_{1,2}(t)|\nu_{1,2} (0) \rangle$ can be obtained from the field theory by considering the time evolution of a single particle state $|1_{\mathbf k}\rangle$. To do that, let us first express the field describing the mass eigenstates using the creation and annihilation operators
\be
\nu_j (\mathbf k,t=0)= \frac{1}{\sqrt{2 E_j}} \, \left(a_{\mathbf k} + a_{-{\mathbf k}}^{\dagger}\right),
\ee
where
\be
\nu_j (\mathbf x,t=0) = \int \frac{d^3 k}{\left(2 \pi\right)^3} \, \nu_j(k) \, e^{i \mathbf k \cdot \mathbf x},
\ee
and $E_j^2 = k^2 + M_j^2$.

The time evolution $\langle \nu_{1,2}(t)|\nu_{1,2} (0) \rangle$ is then given by
\bea
\langle \nu_j (t)|\nu_j (0) \rangle &=& \langle 1_{\mathbf k}| \, e^{-i H t}\, |1_{\mathbf k}  \rangle \nonumber \\
&=& 2 E_j \,  \langle 0| \, \nu_j (\mathbf k,t) \, \nu_j^{\dagger} (\mathbf k,t=0)\,|0  \rangle \nonumber \\
&=& 2 E_j \, \int \frac{dE}{2 \pi i} \, e^{-i E t}\, G_j(E,\mathbf k) \label{timeev}\\
&=& 2 E_j \int \frac{dE}{2 \pi i} \, \frac{e^{-i E t}}{E^2 - k^2 - M_j^2 + i \epsilon} \nonumber \\
&=& e^{-i E_j t}\, .
\eea
Therefore the amplitude is
\bea
|\langle \nu_a(t)|\nu_b(0) \rangle|^2 &=& \sin^2 2\theta \, \sin^2 \frac{\left(E_2-E_1\right) t}{2} \nonumber \\
\Rightarrow |\langle \nu_a(L)|\nu_b(0) \rangle|^2 &\approx&  \sin^2 2\theta \, \sin^2 \frac{\left(M_2^2-M_1^2\right) L}{4 k},
\eea
where on the second line we have used the ultra-relativistic approximation. Here, L is the length of the baseline used in the experiment.

Let us apply this to the $\tilde{2}+1$-scenario. First of all, the overlaps $\langle \nu_{a_{1,2}}|\nu_{1,2,3} \rangle$ can be derived from Eqs. (\ref{0angle}) - (\ref{heavyangle}). Let us start by expressing the energy eigenstates in terms of the ``flavor" eigenstates as
\bea
|\nu_1\rangle_{E,k} &=& - \sin \theta_{E,k} \, |\nu_{a_1}\rangle_k + \cos \theta_{E,k} \, |\nu_{a_2}\rangle_k\,, \\
|\nu_2 \rangle_{E,k} &=& \cos \phi_{E,k} \, \cos \theta_{E,k} \, |\nu_{a_1} \rangle_k + \cos \phi_{E,k} \, \sin \theta_{E,k} \, |\nu_{a_2} \rangle_k - \sin \phi_{E,k} \, |\nu_U \rangle_k \, ,   \\
|\nu_3 \rangle_{E,k} &=& \sin \phi_{E,k} \, \cos \theta_{E,k} \, |\nu_{a_1} \rangle_k + \sin \phi_{E,k} \, \sin \theta_{E,k} \, |\nu_{a_2} \rangle_k + \cos \phi_{E,k} \, |\nu_U \rangle_k \, ,
\eea
where we have made explicit the energy and momentum dependence. Since we are using one-particle quantum mechanics to describe the system, the energy eigenstates must be evaluated on-shell. Therefore
\bea
|\nu_1\rangle &=& - \sin \theta \, |\nu_{a_1}\rangle + \cos \theta \, |\nu_{a_2}\rangle \,, \label{0simp} \\
|\nu_2 \rangle &=& \cos \Phi_2 \, \cos \theta \, |\nu_{a_1} \rangle + \cos \Phi_2 \, \sin \theta \, |\nu_{a_2} \rangle - \sin \Phi_2 \, |\nu_U \rangle \, ,  \label{lightsimp} \\
|\nu_3 \rangle &=& \sin \Phi_3 \, \cos \theta \, |\nu_{a_1} \rangle + \sin \Phi_3 \, \sin \theta \, |\nu_{a_2} \rangle + \cos \Phi_3 \, |\nu_U \rangle \, \label{heavysimp},
\eea
where $\phi_i$ means that the angle $\phi$ is evaluated at $p^2={\cal M}_i^2$, and we have dropped the index for the angle $\theta$ as it is a constant and does not depend on the four-momentum. We note that $\sin \phi_3$ and $\cos \phi_3$ are complex as ${\cal M}_3 > M$. Here, we have also dropped the indices on the states to simplify the notation. We can then solve Eqs. (\ref{0simp}) - (\ref{heavysimp}) for the ``flavor" eigenstates to obtain the overlaps.

Next, we need the time evolution of the mass eigenstates $\langle \nu_i (t)|\nu_i (0) \rangle$, especially for $i=2,3$.  We can obtain these by substituting their respective propagators into Eq. (\ref{timeev}) and it is convenient to use the dispersive form of the propagators \cite{Bjorken:1979dk}. The time evolution is then given by
\bea
S_2 &\equiv& \langle \nu_2 (t)|\nu_2 (0) \rangle \nonumber \\
&=& \int_0^{\infty} dQ^2 \,  \rho_2(Q^2)\,  e^{-i \sqrt{k^2+Q^2} \, t} \nonumber \\
&=& Z_2\,e^{- i E_2 t} + e^{- i E_M t} \int_0^{\infty} dx \, \rho_2(x) \, \exp\left[-i \, \frac{M^2}{E_M} t \, x\right], \label{S2ofti}
\eea
where we have introduced $E_M(k) = \sqrt{k^2 +
{M}^2} $, and $x$ is the dimensionless variable defined in Eq. (\ref{dimvars}).

For the case of $\nu_2$, since the pole is below the continuum, we can estimate the large time behavior of $S_2$ by replacing $\rho_2$ with its near-threshold behavior. Therefore
\bea
S_2 &=& Z_2\,e^{- i E_2(k) t} + \frac{\Delta^2}{4 \pi} \, \sin (2 \pi \eta) \, e^{- i E_M(k) t} \int_0^{\infty} dx \, x^{2 \eta} \, \exp\left[-i \, \frac{M^2}{E_M} t \, x\right] \nonumber \\
&=& Z_2 \, e^{- i E_2 t} + \frac{A_2}{i^{1+2 \eta}}\, \left(\frac{E_M}{M^2 \, t}\right)^{1+2 \eta} \, e^{-i E_M t} \, ,\label{threshcont}
\eea
with
\be
A_2 = \frac{\Delta^2}{4 \pi} \, \sin (2 \pi \eta)  \, \Gamma(1+2\eta) \,. \label{A2coef}
\ee

For $\nu_3$, we can estimate $S_3$ by approximating the spectral density by a Breit-Wigner Lorentzian, (see Eq. (\ref{alfaminbeta})) from which we obtain
\bea
S_3 &=& Z_3 \, e^{- i E_3(k) t}\, e^{- \frac{\Gamma}{2\gamma} t}  ,
\eea
where $\Gamma$ is the decay width given by Eq. (\ref{gama}) and $\gamma = E_3(k)/\mathcal{M}_3$ is the Lorentz factor.

Finally, by combining all the results above, we find that at large time
\bea
|\langle \nu_{a_2}(t)|\nu_{a_1}(0) \rangle|^2 &=& \frac{\sin^2 2\theta}{4} \Bigg\{1 + \cos^4\Phi_2 \left[Z_2^2+ A_2^2 \left(\frac{E_M}{M^2 \, t}\right)^{2+4\eta}\right] - 2 \, Z_2 \, \cos^2 \Phi_2 \, \cos \big[(E_2-E_1)\, t\big] \nonumber \\
& & + \, 2 \,A_2\, Z_3 \, \cos^2 \Phi_2 \, |\sin \Phi_3|^2  \, e^{-\frac{\Gamma}{2\gamma} t} \, \left(\frac{E_M}{M^2 \, t}\right)^{1-\eta} \, \cos \big[(E_3-E_M) t - \frac{\pi}{2}(1+2 \eta)\big] \nonumber \\
& &+ \, Z_3^2 \, |\sin \Phi_3|^4  e^{-   \frac{\Gamma}{\gamma} t} \Bigg\} \label{int} \\
\Rightarrow |\langle \nu_{a_2}(L)|\nu_{a_1}(0) \rangle|^2&\approx& \frac{\sin^2 2\theta}{4} \Bigg\{1 + \cos^4\Phi_2 \left[Z_2^2+ A_2^2 \left(\frac{E_M}{M^2 \, L}\right)^{2+4\eta}\right] - 2 \, Z_2 \, \cos^2 \Phi_2 \, \cos \left[\frac{{\cal M}_2^2\, L}{2 k}\right] \nonumber \\
& & + \, 2 \,A_2\,Z_3 \, \cos^2 \Phi_2 \, |\sin \Phi_3|^2 \, e^{-\frac{\Gamma}{2\gamma} L} \, \left(\frac{E_M}{M^2 \, L}\right)^{1-\eta} \, \cos \left[\frac{({\cal M}_3^2 - M^2)\, L}{2 k} - \frac{\pi}{2}(1+2 \eta)\right] \, \nonumber \\
& &+ \, Z_3^2\, |\sin \Phi_3|^4  e^{-   \frac{\Gamma}{\gamma} L} \Bigg\}\,,  \label{inL}
\eea
where in going from Eq. (\ref{int}) to Eq. (\ref{inL}) we have used the ultra-relativistic approximation and replaced $t$ by the baseline $L$.

\end{document}